# Enhancing Phosphorus Uptake in Sugarcane: A Critical Evaluation of Humic Acid and Phosphorus Fertilizers' Effectiveness

Mandana Mirbakhsh[1]* and Zahra Zahed[2]

[1]Department Agronomy, Purdue University, West Lafayette, IN 47907, USA

[2]Department of Plant Physiology, Alzahra University, Vanak Village Street, Tehran, Iran

*Corresponding Author
Mandana Mirbakhsh, Department Agronomy, Purdue University, West Lafayette, IN 47907, USA.

Submitted: 2023, Jul 15; Accepted: 2023, Aug 14; Published: 2023, Sep 05

## Abstract

Our research conducted in an area characterized by alkaline, lime-abundant soils – investigated the potential of utilizing phosphorus fertilizer and humic acid to enhance phosphorus absorption in sugarcane crops. The results indicated that the application of phosphorus fertilizer significantly increased the total and bioavailable phosphorus in the rhizospheric soil, despite observing a decrease in phosphatase enzyme activity.

An important observation was the considerable growth of active carbon, a crucial soil health indicator, under the influence of humic acid treatments. The findings also demonstrated an enhancement in phosphorus absorption by sugarcane due to the synergistic application of humic acid and phosphorus fertilizer at both harvest periods. Interestingly, humic acid treatments, when applied through immersion, were found to be more effective than soil applications, implying a greater impact on root absorption processes. The findings underline the potential of integrating humic acid into sugarcane cultivation for better phosphorus absorption.

Our study offers valuable insights for improved soil management strategies, and could potentially pave the way towards more sustainable agricultural practices. However, we also recommend further investigation into alternative methods of humic acid application and its usage at different stages of plant growth. Such exploration could provide a comprehensive understanding of the potential benefits and most effective utilization of humic acid in agriculture, especially in regions with similar soil characteristics as West Azarbaijan, Iran.

## 1. Introduction

The cultivated sugarcane soils located in West Azarbaijan, western Iran, are characterized by limited organic matter (less than 0.5%), as well as high pH (8-9), and rich in lime (above 25%). Moreover, issues related to drainage, salinity, and alkalinity have been observed in many of these regions [1]. Such conditions, particularly the high lime content and pH range, create challenges concerning phosphorus mobility in these soils, subsequently affecting the plant's ability to absorb phosphorus. These complications have been well documented in previous studies. Given this context and considering the long history of sugarcane cultivation—spanning over 50 years in the north and 30 years in the west of Azarbaijan province the continued application of phosphorus fertilizer appears necessary despite its longstanding usage.

Currently, the addition of phosphorus fertilizer occurs only during the planting stage, with no subsequent fertilization in the re-growth phases. This application is performed at a rate of 250-300 kilograms per hectare. Recent years have witnessed growing global interest in improving the organic matter content of soils, employing strategies such as compost application and utilizing by-products of sugar factories, notably filter cake [2]. Accompanying research has suggested that organic compounds can enhance the mobility of certain nutrients, especially phosphorus, and facilitate their absorption by various plants, including sugarcane [3]. The results gleaned from these investigations point towards a positive impact on sugarcane growth and performance, as well as an enhancement in phosphorus absorption and the efficiency of its consumption [4-7].

The procurement of requisite raw organic materials, such as farm by-products, incurs substantial costs. Additionally, the time investment necessary to realize their impact on soil improvement can be significant. Due to these constraints, alternative approaches such as the use of extracts, low volume or compressed derivations of these substances—including composts, biochars, and humic acid, nano fertilizer—have been explored [8,9].

Humic acid, characterized by its organic structure, possesses properties capable of influencing the mobility and availability of nutrients within the soil [10,11]. Notably, humic acid has been observed to augment nutrient uptake, such as phosphorus, by fostering root growth and development, thereby potentially enhancing overall plant growth and performance [12]. Improving nutrient absorption via root expansion, as well as enhancing the length and density of root hairs, are among the most impactful effects of humic acid [13]. Several researchers have highlighted the role of humic acid in augmenting the soil microbial population, intensifying the secretion of organic acids in the rhizosphere, thereby advancing the mobility and absorption of nutrients [14-18].

Soil enzymes also have a crucial function in the decomposition of organic materials and in the nutrient cycling [19,20]. Within the process of nutrient absorption, specific soil enzymes have been found to amplify the rate of reactions involved in the decomposition of plant residues and the subsequent release of plant-utilizable nutrients. These soil enzymes originate from three primary sources: 1) living and non-living microorganisms, 2) roots and plant residues, and 3) soil organisms [21,22].

Phosphatase is a pivotal soil enzyme responsible for the conversion of organic phosphorus into mineral phosphorus, which is plant-available, thus acting as a potential indicator of biological alterations within the soil [23,24]. Enzymes such as alkaline and acidic phosphatases demonstrate enhanced activity in alkaline and acidic soils respectively, contingent upon the soil's pH levels. The availability of organic material sources and phosphorus are key factors influencing these enzymes' activity [25-28].

Active portions of soil carbon, often termed as 'active carbon reserves,' are recognized as sensitive indicators integral to soil quality management [29,30]. This segment of soil carbon holds significant importance for the soil food web, thereby impacting nutrient cycles and biological properties of the soil. In essence, the decomposable component of organic carbon represents an active element with a comparably high turnover rate. Its volume exhibits correlation with the soil microbial biomass, thus rendering it a key factor in the potential fertility of the soil [31,32]. Some researchers utilize active carbon as a yardstick to denote soil quality [33,34].

The primary objective of this research is to explore the extent of phosphorus absorption by sugarcane plants, attributed to modifications in the plant rhizosphere encompassing phosphatase enzyme activity, active carbon, and accessible rhizosphere soil phosphorus. This is to be examined under the influence of phosphorus fertilizer and humic acid treatments, across two distinct harvest periods.

## 2. Material and Methods

In order to compare the effects of phosphorus fertilizer and humic acid on phosphatase enzymes (alkaline and acidic), active carbon (labile carbon), and soil phosphorus in the sugarcane rhizosphere (total phosphorus and Olsen phosphorus), a pot experiment was conducted in a greenhouse in western Iran, at the Hakim Farabi Agro-Industry Company (48º36' E, 30º59' N). In this experiment, different levels of phosphorus (0, 50, and 100 percent of fertilizer recommendation equivalent to 250 kilograms per hectare), applied as soil amendments and placed beneath the sugarcane cuttings, treatments of humic acid (three levels of soaking cuttings in 0, 0.3, and 0.5 percent humic acid solutions, and soil application at the rate of 10 kilograms per hectare of humic acid), and two harvest times (45 and 90 days after planting) were investigated. In the soaking treatments, after removing the husk, sugarcane cuttings were soaked for 30 minutes in humic acid solutions at the mentioned concentrations.

### 2.1. Preparing the Pots, Soil Analysis, and Humic Acid

The soil used in this study was collected from the surface layer of the soil (0-30 centimeters) of a sugarcane farm located in the Hakim Farabi Agro-Industry Company. The soil samples were air-dried, crushed, passed through a 2-millimeter sieve, and placed in pots with a diameter of 30 centimeters and a depth of 50 centimeters. To protect the roots and ensure that they were not damaged during plant removal, appropriately sized plastic bags were placed in the pots. The physical and chemical properties of the soil sample were determined using standard laboratory methods (Table 1). Soil texture was determined by the hydrometer method electrical conductivity according to the Rhoades method (1996), and calcium carbonate equivalent was measured by titration with acid [35,36].

| Soil texture | Clay (%) | Silt (%) | Sand (%) | Carbon (%) | Phosphorus (mg/kg) | $CaCo_3$ (%) | pH | EC (dSm$^{-1}$) |
|---|---|---|---|---|---|---|---|---|
| Silt-loam | 14 | 21 | 65 | 0.18 | 3.6 | 32.5 | 8.03 | 3.52 |

**Table 1: Physical and Chemical Properties of Soil**

The soil pH was determined in a 1:2 soil-to-water suspension. The soil organic carbon was determined using the wet oxidation method (Nelson and Sommers, 1996), and phosphorus extracted with sodium bicarbonate (Olsen phosphorus) was determined by the [37]. The humic acid extracted from filter cake was analyzed using elemental analysis (Costech ECS 4010 CHNSO model) to determine carbon, nitrogen, sulfur, and hydrogen. The amounts of these elements were determined to be 12.53%, 2.52%, 1.2%, and 0.2%, respectively [38].

### 2.2. Cultivation and Execution of the Experiment
For planting, the soil was mixed with sieved and washed sand at a ratio of 1:1, and the pots were filled. Subsequently, single-bud cuttings of the 69CP-1062 variety were planted in the first week of October 2016. Before planting, to perform the soaking treatment, the cuttings were soaked in 0.3% and 0.5% humic acid solutions and a control treatment in water for 30 minutes. Phosphorus fertilizer was also applied at levels of zero, 50%, and 100% of the recommended amount for the region before planting. Nitrogen fertilizer (urea) was applied in two stages during the plant's growth period at a rate equivalent to 50 kilograms per hectare for all treatments.

To maintain soil moisture within 80% of the field capacity, irrigation was performed every two days. The soil moisture level before initiating irrigation was determined by measuring the moisture content in destructive pots.

### 2.3. Harvesting
The sugarcane plant was harvested at the connection point to the soil and the collar of the plant 45 and 90 days after planting and was placed in a plastic bag for later tests. The pot soil, along with the plastic covering inside the pot, was carefully removed. After removing the sugarcane roots, the soil adhering to the root (rhizosphere soil) was carefully and gently separated and collected.

### 2.4. Rhizosphere Soil Analyses
The active (labile) carbon in the soil was determined in accordance with the Weil et al. method (2003), total phosphorus was determined by the and phosphorus extracted with sodium bicarbonate (Olsen phosphorus) was determined by the Olsen et al. method (1954). Additionally, the activities of alkaline and acidic phosphatase enzymes were measured using the [39,40].

### 2.5. Absorption Calculation
To determine the dry weight of the aerial part and the root of the plant, the samples were placed in an oven at 70 degrees Celsius for 48 hours. Phosphorus in the dried plant samples was determined after grinding, using the [41]. After determining the dry weight and phosphorus content of the plant, the amount of phosphorus absorbed was calculated from the product of phosphorus concentration and dry weight of the plant.

### 2.6. Statistical Analysis
In this study, a factorial experiment was conducted using two factors, phosphorus fertilizer and humic acid, with three replications in a completely randomized design. The statistical calculations were performed using R-studio 4.2.2 software, and the comparison of means was conducted using Tukey's test at the 5% level.

### 3. Results
The results of the analysis of variance of the mean squares of different treatments are presented in Tables 2 and 3. As these tables show, the application of humic acid and phosphorus fertilizer resulted in significant changes (at 1% and 5% levels) in all the factors under investigation, except for acidic phosphatase at one or both harvest times.

| S. O. V | df | Total P (mg kg$^{-1}$) | Olsen P (mg kg$^{-1}$) | Alkaline Phosphatase (µmol h$^{-1}$ g$^{-1}$) | Acidic Phosphatase (µmol h$^{-1}$ g$^{-1}$) | Active Carbon (mg kg$^{-1}$) | P absorption (mg P plant$^{-1}$) |
|---|---|---|---|---|---|---|---|
| Phosphorus fertilizer (A) | 2 | 24047** | 136.69** | 8.07* | 1.18$^{ns}$ | 411.70$^{ns}$ | 26.72* |
| Humic acid (B) | 3 | 10426$^{ns}$ | 3.88$^{ns}$ | 1.84$^{ns}$ | 0.1256$^{ns}$ | 82203.3** | 63.53** |
| (A)×(B) | 6 | 4663$^{ns}$ | 2.88$^{ns}$ | 0.75$^{ns}$ | 19.74$^{ns}$ | 772.8$^{ns}$ | 2.42$^{ns}$ |
| Error | 24 | 4433 | 2.36 | 1.26 | 44.83 | 769.60 | 2.87 |

Non-significant differences (ns), significant difference at 5% (*), significant difference at 1% (**)

**Table 2:** Variance decomposition of the effect of phosphorus fertilizer and humic acid on phosphorus, active carbon, phosphatase enzymes in the rhizosphere soil, and absorption of phosphorus by sugarcane plant 45 days after planting

| S. O. V | df | Total P (mg kg$^{-1}$) | Olsen P (mg kg$^{-1}$) | Alkaline Phosphatase (µmol h$^{-1}$ g$^{-1}$) | Acidic Phosphatase (µmol h$^{-1}$ g$^{-1}$) | Active Carbon (mg kg$^{-1}$) | P absorption (mg P plant$^{-1}$) |
|---|---|---|---|---|---|---|---|
| Phosphorus fertilizer (A) | 2 | 15652.9** | 50.11** | 2.37* | 0.15ns | 2032ns | 120.85* |
| Humic acid (B) | 3 | 9.803ns | 1.36ns | 0.9853ns | 0.4251ns | 7621.6** | 210.00** |
| (A)×(B) | 6 | 832.1ns | 2.75ns | 0.5108ns | 0.766ns | 1074.4ns | 12.76ns |
| Error | 24 | 510.4 | 1.26 | 0.8182 | 0.2091 | 961.9 | 14.25 |

Non-significant differences (ns), significant difference at 5% (*), significant difference at 1% (**)

**Table 3:** Variance decomposition of the effect of phosphorus fertilizer and humic acid on phosphorus, active carbon, phosphatase enzymes in the rhizosphere soil, and absorption of phosphorus by sugarcane plant 90 days after planting.

### 4. Soil Phosphorus
### 4.1. Total Phosphorus

As delineated in the analysis of variance table, total phosphorus exhibited a statistically significant augmentation at a confidence level of 1% as a consequence of the application of phosphorus fertilizer for both harvest times. Changes induced by the application of humic acid treatments did not reach statistical significance at either of the harvest periods.

The data presented in Tables 4 and 5 illustrate the comparative mean total phosphorus in the rhizosphere soil, indicating that the observed fluctuations were principally correlated with the applied levels of phosphorus fertilizer.

This pattern was congruent in both harvest periods, suggesting that the increment in phosphorus fertilizer application was significantly associated with an increase in total rhizosphere soil phosphorus content. During the first harvest, the minimal quantities of total phosphorus were associated with treatments devoid of phosphorus fertilizer application. Conversely, the maximal quantities were associated with treatments involving phosphorus fertilizer application, specifically at 100% of the recommended fertilizer application rate. In the majority of the fertilizer treatments, the deployment of humic acid demonstrated higher total phosphorus levels as compared to its absence. A similar pattern emerged during the second harvest, albeit with a more distinct divergence between treatments. In this interval, the highest total phosphorus content corresponded to phosphorus fertilizer treatments at 100% of the recommended application rate, combined with the application of humic acid. The smallest phosphorus content was associated with treatments abstaining from phosphorus fertilizer application (as depicted in Tables 4 and 5).

In terms of numerical values, the total phosphorus content during the first harvest surpassed that of the second harvest. Moreover, in the treatments involving humic acid application to the soil, a more pronounced difference in total phosphorus content was observed. During the second harvest, notwithstanding the decrease in total phosphorus content, its level consistently ascended with the escalation in fertilization.

### 4.2. Bioavailable Phosphorus (Olsen P)

As demonstrated in Tables 2 and 3, the bioavailable phosphorus (Olsen P) in rhizospheric soil exhibited a significant enhancement at a 1% confidence level in both harvest periods due to the application of phosphorus fertilizer. Conversely, the employment of humic acid treatments did not instigate a meaningful variation in bioavailable phosphorus. The trajectory of bioavailable phosphorus during both harvest durations exhibited a similar pattern. This suggested that its fluctuations were predominantly contingent upon the phosphorus fertilizer treatments, a pattern that mirrored that of total phosphorus. The individual effects of the treatments clearly exhibited the relationship between changes in Olsen P and the increased use of phosphorus fertilizer. Interestingly, in the interplay of treatments, akin to total phosphorus, the combined application of humic acid and phosphorus fertilizer showcased the highest levels of bioavailable phosphorus.

As delineated by the results in Tables 4 and 5, the quantity of bioavailable phosphorus in the initial harvest exceeded that in the subsequent harvest. Additionally, during both periods, its concentration escalated with the intensification of phosphorus fertilizer application. The highest concentration of bioavailable phosphorus during the first and second harvests corresponded to treatments that incorporated humic acid application (both stem immersion and soil application) along with 100% recommended rate of phosphorus fertilizer application. The minimal concentration of bioavailable phosphorus in both harvests was associated with treatments refraining from phosphorus fertilizer application. The correlation between total phosphorus and Olsen P also manifested a significant association during both harvests. Nevertheless, this relationship during the first harvest was associated with a superior correlation coefficient (as seen in Tables 4 and 5, Figures 1 and 2.

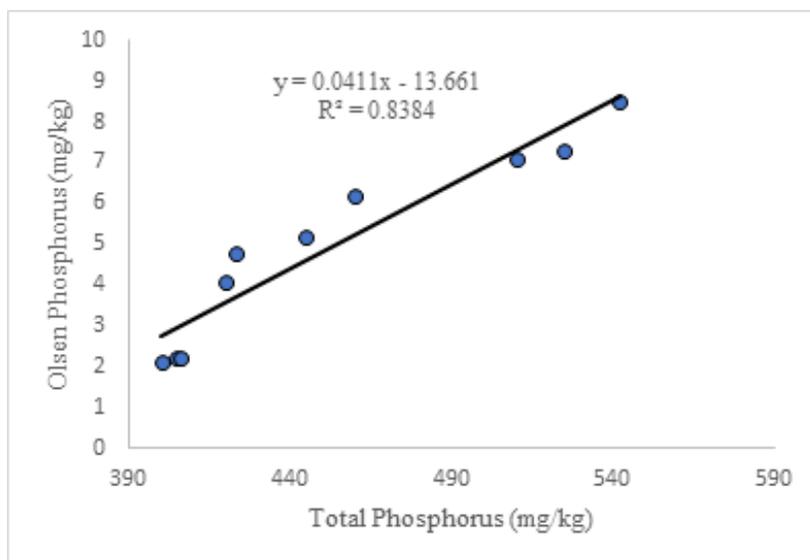

**Figure 1:** Olsen and Total Phosphorus in the First Harvest

### 4.3. Alkaline Phosphatase Enzyme
The application of various humic acid treatments did not result in significant changes in the activity of the alkaline phosphatase enzyme. This occurred even though the use of phosphate fertilizer significantly reduced the amount of this enzyme in the rhizosphere soil at the 5% level at both harvest times (Tables 2 and 3). These results suggest that phosphate fertilizer application had an inverse effect on alkaline phosphatase activity during both harvest times.

Comparing the means of different treatments indicates that, contrary to phosphate fertilizer, humic acid application in treatments without fertilization and with low fertilizer levels (50% recommended fertilizer application) showed higher phosphatase levels in both harvest times. The alkaline phosphatase enzyme decreased over time in most treatments during both harvest times (Tables 4 and 5).

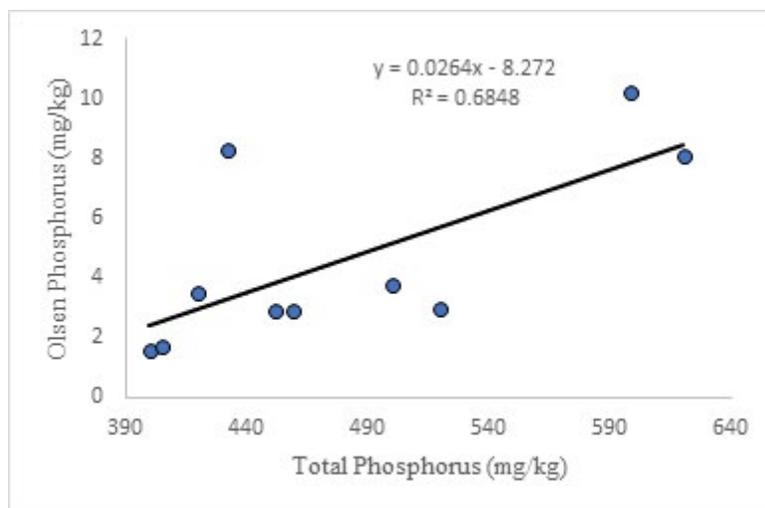

**Figure 2:** Olsen and Total Phosphorus in the Second Harvest

Although in both harvest times, the highest and lowest levels of alkaline phosphatase were related to the treatment with humic acid application and the treatment with maximum phosphate fertilizer application (8.51 and 7.09 micromoles per hour per gram, respectively, in the first and second harvests), and the treatment without phosphate fertilizer application, regardless of the use of humic acid (5.19 micromoles per hour per gram).

### 4.4. Acid Phosphatase Enzyme
As the analysis of variance table indicates, the reaction of acid phosphatase to both phosphorus fertilizer and humic acid treatments was not statistically significant at either harvest time (Tables 2 and 3). Although the mean comparison table does not indicate a significant effect for either phosphorus or humic acid treatments, the trend of changes in the activity of this enzyme in the first harvest was similar to alkaline phosphatase. In this period, the lowest amount of acid phosphatase was related to phosphorus fertilizer treatments (100% of the recommended fertilizer). However, it did not show a clear trend in comparison with the first harvest. The amount of acid phosphatase in different treatments, like alkaline phosphatase, was higher in the first harvest than in the second (Tables 4 and 5). In the first harvest, humic acid in immersion treatments showed higher acid phosphatase than soil treatments in most cases. Examining the relationship between phosphatase enzymes and soil phosphorus (total phosphorus and available phosphorus) in the first and second harvest showed noticeable variations. Total phosphorus showed an inverse trend with both enzymes at both harvest times, but the correlation coefficient was higher for alkaline phosphatase. The same trend was observed for available phosphorus (Figures 3-6).

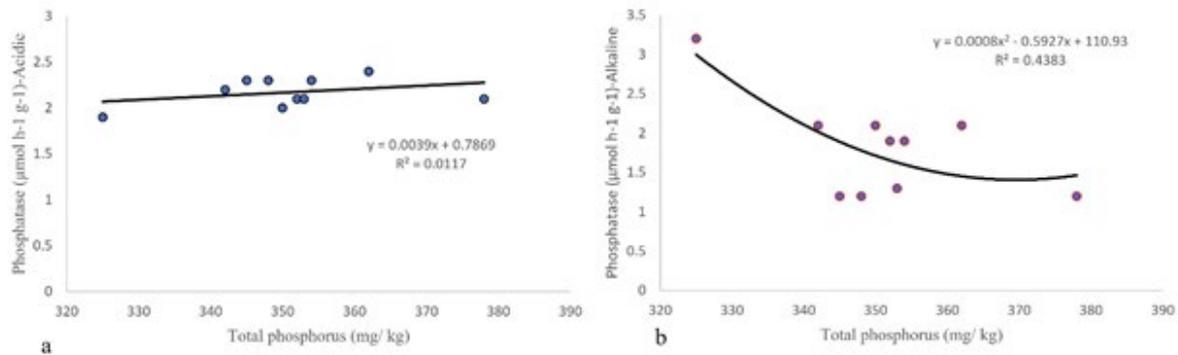

**Figure 3:** The relationship between total phosphorus and phosphatase enzymes in the first harvest. (a) Acidic phosphatase, (b) Alkaline phosphatase

### 4.5. Active Carbon
The results from the analysis of variance table indicate that the usage of humic acid treatments significantly increased the quantity of active carbon in the rhizosphere soil at a level of 1 percent (Table 2 and 3).

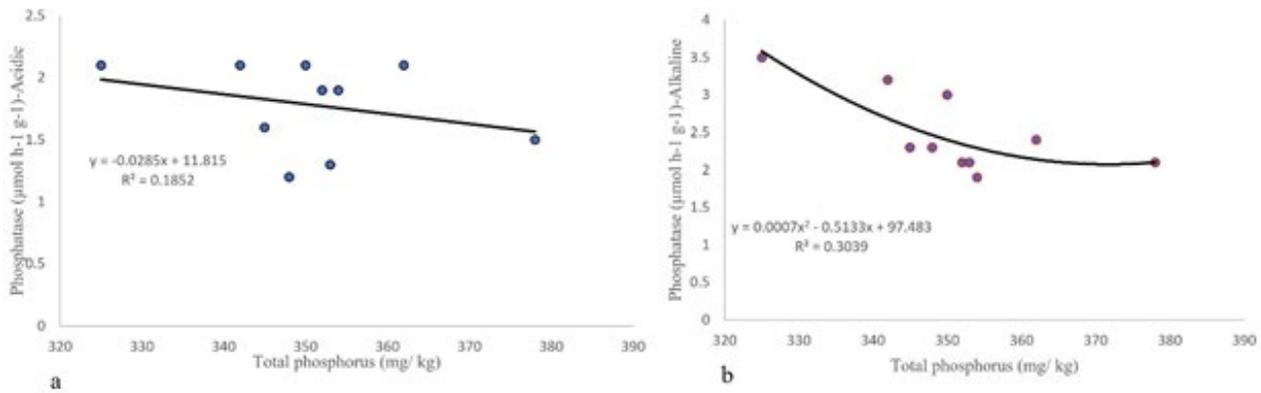

**Figure 4:** The relationship between total phosphorus and phosphatase enzymes in the second harvest. (a) Acidic phosphatase, (b) Alkaline phosphatase

The comparison of the average amount of active carbon in different treatments at two harvest times showed a completely different trend compared to alkaline and acidic phosphatase, but similar to total phosphorus and available phosphorus. Contrary to the phosphatase enzymes, the amount of active carbon at the second harvest time was higher in most treatments than at the first harvest. However, like the phosphatase enzymes, the humic acid treatments, both in the form of cutting dip and soil application, showed a higher quantity of active carbon. The amount of active carbon at both harvest times, compared to phosphorus fertilizer, had a greater dependency on humic acid, but this dependency was more pronounced at the first harvest. The effect of phosphorus alone, although not significant in both harvests, increased the amount of active carbon with an increased use of phosphorus fertilizer. In the first harvest, the highest and lowest quantities of active carbon, with values of 264 and 87.181 mg/kg respectively, were related to the 100P0.3H and 0P0H treatments. In the second harvest, similar to the first harvest, humic acid treatments, compared to non-use, showed a greater amount of active carbon (Tables 4 and 5).

| Treatments | Total P (mg kg$^{-1}$) | Olsen P (mg kg$^{-1}$) | Alkaline Phosphatase (μmol h$^{-1}$ g$^{-1}$) | Acidic Phosphatase (μmol h$^{-1}$ g$^{-1}$) | Active Carbon (mg kg$^{-1}$) | P absorption (mg P plant$^{-1}$) |
|---|---|---|---|---|---|---|
| $H_0P_0$ | 404.33 ab | 2.80 d | 6.84 ab | 2.87 a | 181.87 b | 5.41 e |
| $H_0P5_0$ | 419.17 ab | 5.67 cd | 6.12 ab | 3.14 a | 184.46 ab | 5.73 de |
| $H_0P_{100}$ | 459.00 ab | 7.50 abc | 5.19 b | 3.07 a | 185.98 ab | 8.45 be |
| $H_{0.3}P_0$ | 394.5 b | 2.53 d | 8.51 a | 3.02 a | 220.00 ab | 10.41 ad |
| $H_{0.3}P_{50}$ | 446.50 ab | 6.27 bcd | 6.30 ab | 3.02 a | 249.10 ab | 10.29 ae |
| $H_{0.3}P_{100}$ | 467.56 ab | 8.50 abc | 5.94 ab | 2.73 a | 264.00 a | 14.79 a |
| $H_{0.5}P_0$ | 392.5 b | 2.53 d | 6.40 ab | 3.05 a | 219.62 ab | 10.11 ae |
| $H_{0.5}P_{50}$ | 450.50 ab | 5.47 cd | 6.01 ab | 3.05 a | 212.33 ab | 11.11 abc |
| $H_{0.5}P_{100}$ | 556.42 ab | 10.30 ab | 5.46 ab | 2.56 a | 244.69 ab | 12.22 ab |
| $H_{SA}P_0$ | 387.50 b | 2.67 d | 7.09 ab | 2.95 a | 260.82 ab | 7.08 cde |
| $H_{SA}P_{50}$ | 529.67 ab | 6.73 ad | 6.95 ab | 3.39 a | 243.29 ab | 6.97 cde |
| $H_{SA}P_{100}$ | 590.92 a | 11.23 a | 5.69 ab | 2.39 a | 261.20 ab | 8.57 be |

$H_0$, $H_{0.3}$, $H_{0.5}$, refer to the solutions of zero, 0.3%, and 0.5% humic acid, respectively. $P_0$, $P_{50}$, $P_{100}$ refer to the solutions of zero, 50% and 100% humic acid, respectively. HSA, indicates the soil application of humic acid. (The average of the numbers with similar letters in each column is not significant according to Tukey's test at the 5% level).

**Table 4:** Comparison of averages between active carbon, phosphatase enzymes, rhizospheric soil phosphorus, and phosphorus absorption of sugarcane in the first harvest.

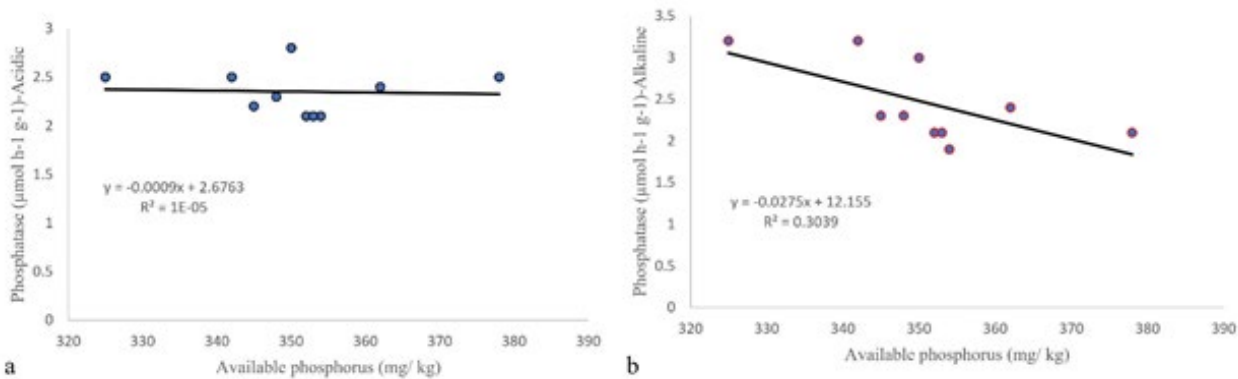

**Figure 5:** The relationship between Olsen phosphorus (available phosphorus) and phosphatase enzymes in the first harvest. (a) Acidic phosphatase, (b) Alkaline phosphatase

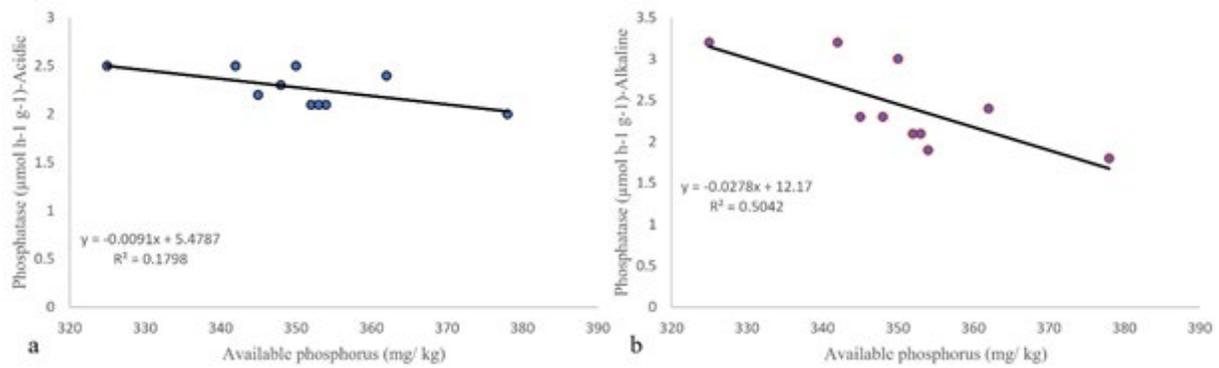

**Figure 6:** The relationship between Olsen phosphorus (available phosphorus) and phosphatase enzymes in the first harvest. (a) Acidic phosphatase, (b) Alkaline phosphatase

### 4.6. Phosphorus Absorption

The absorption of phosphorus by sugarcane was significantly increased by the application of both humic acid and phosphorus fertilizer at both harvest times, with these changes significant at the 1% level for humic acid application and at the 5% level for phosphorus fertilizer application (Tables 2 and 3). The comparison of average phosphorus absorption in different treatments and at both harvest times shows that phosphorus absorption was significantly higher in the treatments with humic acid at both harvest times (Tables 4 and 5). Phosphorus absorption in various treatments increased significantly with the addition of phosphorus and humic acid (either individually or in combination with phosphorus application) compared to the control treatment, and in most cases, this was significant at the 5% level. The highest and lowest phosphorus absorption at the first and second harvest were respectively related to the treatment of dipping cuttings in a 0.3% solution along with phosphorus fertilizer application (100% fertilizer recommendation) and the control treatment. Although the treatments with humic acid application to the soil showed lower absorption compared to other humic acid treatments, they still had significantly higher absorption compared to the control treatment. The amount of absorption in the second harvest increased significantly compared to the first harvest, such that in different treatments, the absorption in the second harvest was at least three times higher than the first harvest (Table 5). Among the factors under study (phosphatase enzymes, soil phosphorus, and active carbon), only the active carbon factor showed a significant relationship with phosphorus absorption. The correlation coefficient of this relationship increased significantly when soil application treatments were excluded, thus demonstrating a very high correlation coefficient between this factor and phosphorus absorption at both harvest times (Figures 7 and 8).

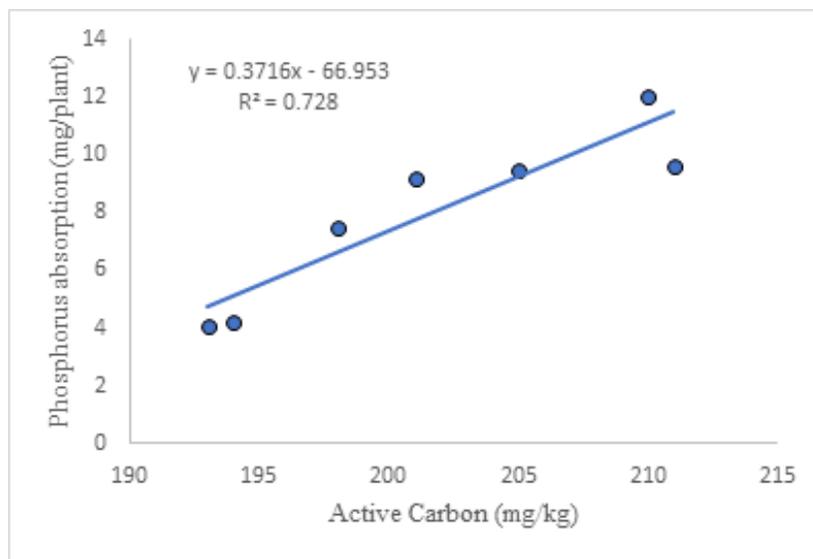

**Figure 7:** Active Carbon and Phosphorus Absorption in the First Harvest

| Treatments | Total P (mg kg$^{-1}$) | Olsen P (mg kg$^{-1}$) | Alkaline Phosphatase (μmol h$^{-1}$ g$^{-1}$) | Acidic Phosphatase (μmol h$^{-1}$ g$^{-1}$) | Active Carbon (mg kg$^{-1}$) | P absorption (mg P plant$^{-1}$) |
|---|---|---|---|---|---|---|
| $H_0P_0$ | 368.00 c | 2.80 cd | 6.25 a | 1.77 a | 179.72 b | 27.69 c |
| $H_0P_{50}$ | 397.00 bc | 3.40 cd | 5.39 a | 1.93 a | 191.98 b | 31.46 bc |
| $H_0P_{100}$ | 413.78 abc | 4.87 ad | 5.37 a | 1.90 a | 219.24 ab | 35.39 bc |
| $H_{0.3}P_0$ | 379.67 c | 1.93 cd | 7.09 a | 2.27 a | 208.66 ab | 36.38 abc |
| $H_{0.3}P_{50}$ | 406.17 bc | 4.00 cd | 6.30 a | 2.39 a | 226.93 ab | 42.47 ab |
| $H_{0.3}P_{100}$ | 426.83 abc | 7.02 ab | 5.19 a | 1.94 a | 260.82 ab | 47.31 a |
| $H_{0.5}P_0$ | 362.33 c | 1.80 d | 6.00 a | 1.92 a | 218.62 ab | 37.52 abc |
| $H_{0.5}P_{50}$ | 411.83 abc | 3.70 cd | 5.63 a | 1.74 a | 256.28 ab | 42.13 ab |
| $H_{0.5}P_{100}$ | 473.00 a | 5.13 abc | 5.26 a | 1.83 a | 224.78 ab | 42.33 ab |
| $H_{SA}P_0$ | 363.50 c | 2.07 cd | 5.41 a | 2.30 a | 263.47 ab | 34.59 bc |
| $H_{SA}P_{50}$ | 391.00 bc | 3.00 cd | 5.43 a | 2.12 a | 285.39 a | 36.05 bc |
| $H_{SA}P_{100}$ | 448.50 ab | 7.66 a | 5.39 a | 2.31 a | 255.91 ab | 36.24 bc |

$H_0$, $H_{0.3}$, $H_{0.5}$, refer to the solutions of zero, 0.3%, and 0.5% humic acid, respectively. $P_0$, $P_{50}$, $P_{100}$ refer to the solutions of zero, 50% and 100% humic acid, respectively. HSA, indicates the soil application of humic acid. (The average of the numbers with similar letters in each column is not significant according to Tukey's test at the 5% level).

**Table 5:** Comparison of averages between active carbon, phosphatase enzymes, rhizospheric soil phosphorus, and phosphorus absorption of sugarcane in the second harvest.

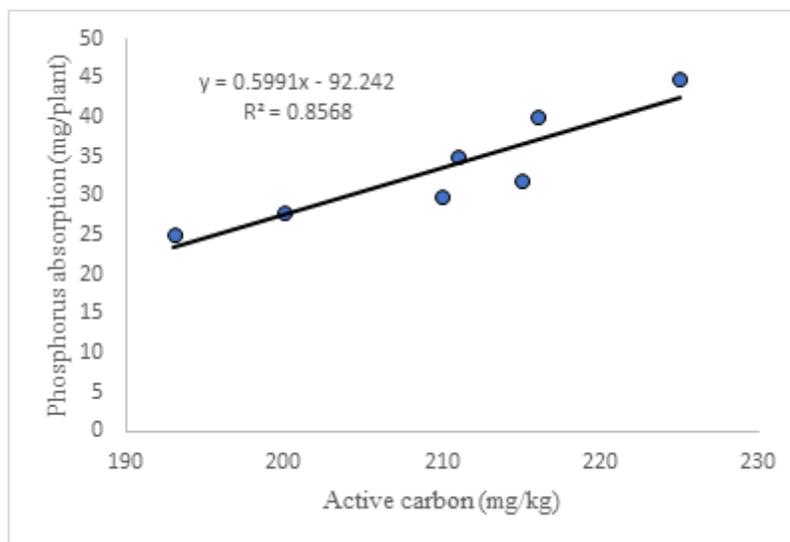

**Figure 8:** Active Carbon and Phosphorus Absorption in the Second Harvest

## 5. Discussion

The obtained results related to the harvest time and fertilizer treatments and humic acid highlight the distinctly significant role of fertilization in increasing total phosphorus and available phosphorus in the rhizosphere soil. The increase in phosphorus with increased phosphorus fertilizer is an obvious result that has been emphasized by many researchers [42]. However, other points have also been clarified in this research. Among them is the significant correlation between total phosphorus and available phosphorus in both harvests, albeit with a considerable difference in the first harvest, as well as a decrease in the amount of phosphorus (both available and total) over time. The reason for these changes can be attributed to changes in phosphorus absorption and its increased amount in the second harvest. With the increased use of phosphorus fertilizer in the 50% and 100% fertilizer recommendation treatments, under the combined effect of humic acid with phosphorus fertilizer compared to treatments without humic acid, the amount of total phosphorus and available phosphorus experiences a larger increase. These results are consistent with the findings of [43].

With increased phosphorus absorption and the impact of various treatments, the relationship between total phosphorus and available phosphorus in the second harvest has been more influenced and these changes, although causing a decrease in data correlation, have not affected the overall result and the significance of their relationship. Changes in soil phosphorus caused changes in enzyme levels, particularly alkaline phosphatase. In accordance with the findings of the amount of alkaline phosphatase was higher than acid phosphatase in both harvest times, due to favorable conditions in soils with a high pH [44].

With the increased use of phosphorus and consequently the increase in total phosphorus and available phosphorus in the rhizosphere soil, the activity of phosphatase enzymes, particularly alkaline phosphatase, decreased. This can be due to the availability of phosphorus in the plant root zone and the lack of need for phosphatase activity to provide the plant's required phosphorus from soil's organic phosphorus sources, as well as the inhibitory or reductive role of phosphorus in the secretion of this enzyme by soil organisms and plants. This aligns with research. This phenomenon is clearly observed with an increase in phosphatase enzyme levels in the absence of fertilizer use. The role of humic acid in this regard, due to the much stronger role of phosphorus, is less observed. However, it seems that the use of humic acid in fertilizer treatments has been able to increase the activity of the phosphatase enzyme (although it was not statistically significant). This could be related to the increased soil microbial population and improved root growth of the plant, as well as an increase in the secretion of this enzyme. Unlike the phosphatase enzyme, active carbon showed higher amounts in the second harvest compared to the first, and in addition to this, the use of phosphorus fertilizer resulted in an increase in the amount of active carbon, as observed [45-47].

In the majority of phosphorus fertilizer treatments, the quantity of active carbon did not demonstrate a significant discrepancy in the initial harvest. However, an augmentation in active carbon was noted in the subsequent harvest due to the utilization of phosphorus fertilizer. Conversely, treatments involving humic acid, especially the ones involving humic acid soil application, demonstrated an increment in active carbon in both harvests. This increase was substantially more pronounced in the second harvest. The rise in active carbon during the second harvest, corresponding with phosphorus fertilizer, can be attributed to the roles played by humic acid, organic substances, and particularly the soil treatment application. These elements contribute to an increase in the soil's organic carbon, prolonged root growth and expansion compared to the initial harvest, and a consequent increase in carbonaceous materials within the soil's growth environment. The more substantial impact of the humic acid soil application compared to immersion treatments could potentially be ascribed to higher consumption of humic acid and its more direct interaction with the soil. With the increase in active soil carbon, an enhancement in phosphorus absorption was noted across both harvests. However, the correlation of absorption with active carbon in humic acid immersion treatments and phosphorus treatments was elevated when compared to data inclusive of humic acid soil application treatments.

More specifically, the correlation of phosphorus absorption with active carbon at both harvest times saw a significant rise with the exclusion of data pertaining to the humic acid soil treatment application. This was reflected by the increase in $r^2$ values from 0.02 to 0.73 during the first harvest and 0.1 to 0.86 in the second harvest. This observation could be reasoned by the increased contact of humic acid with the root zone in immersion treatments. Consequently, it may have resulted in more effective participation in root absorption processes and rhizosphere region compared to soil application treatments. The more significant influence of immersion treatments, in conjunction with phosphorus fertilizer utilization due to a more extended period for root growth improvement, may have resulted in the absorption of more phosphorus.

The overall role of humic acid as an organic compound, alongside phosphorus fertilizer in escalating plant biomass production, could potentially explain the heightened phosphorus absorption. This hypothesis aligns with the findings of [48,49].

The absorption of phosphorus demonstrated a comparable pattern across both harvesting periods, with an observed escalation in absorption commensurate with the increase in phosphorus fertilizer application. This observation aligns with the research findings presented by a temporal absorption comparison between the first and second harvest cycles revealed that despite a reduction in phosphorus concentration, total absorption experienced an increase, which can be attributed to the rise in dry matter. Besides phosphorus consumption, the application of humic acid significantly elevated phosphorus absorption. Interestingly, improvements in phosphorus absorption were observed even in treatments devoid of phosphorus application, due to the employment of humic acid. Analogous results have been reported for alternative crops, such as wheat. Increased phosphorus absorption in treatments involving humic acid may be ascribed to enhancements in the root system, as evidenced by an increase in root length and volume, as well as an augmentation in the length and density of root hairs. Consequently, the application of humic acid appears to bolster the capacity for phosphorus absorption from soil and fertilizer sources by facilitating increases in root length and volume. These findings are congruent with the hypothesis that humic acid and organic compounds can augment the utilization of phosphorus by plants [51-54].

## 6. Conclusion
In the current investigation, diverse treatment modalities involving humic acid and phosphorus fertilizer were scrutinized. Findings reveal an enhancement in the efficiency of phosphorus fertilizer utilization concomitant with the application of humic acid. This enhancement is particularly pronounced in immersion treatments where sugarcane cuttings are placed in a humic acid solution.

Given the observed improvement in phosphorus absorption by the plant, a subsequent reduction in fertilizer usage is anticipated. Consequently, these results suggest that through superior soil management and the creation of conducive conditions for phosphorus mobility in soils designated for sugarcane cultivation (via the application of organic compounds such as humic acid), there is potential to facilitate greater phosphorus absorption. This can be achieved even with diminished utilization of phosphorus fertilizers, as compared to conventional conditions [55-58]. The empirical evidence derived from this study posits that the operational application of humic acid during the cultivation of sugarcane cuttings is plausible. Nevertheless, it appears that additional research is warranted to examine other pertinent considerations, such as the utilization of humic acid during the plant growth season and the exploration of alternate application methods, including foliar spraying or its use in irrigation water.

## Author Contribution

The authors confirm contribution to the paper as follows: study conception and design: Zahra Zahed. data collection: Zahra Zahed. Analysis and interpretation of results: Mandana Mirbakhsh.; draft manuscript preparation: Mandana Mirbakhsh. Revising: Mandana Mirbakhsh, and both authors reviewed the results and approved the final version of the manuscript.

## Conflict of Interest

The authors declare that there is no conflict of interest for our research titled: "Enhancing Phosphorus Uptake in Sugarcane: A Critical Evaluation of Humic Acid and Phosphorus Fertilizers' Effectiveness.", and authors take full responsibility of it.

## Ethical Approval

This study was approved by administrative committee of Research Farm (Islamic Azad University, Azerbayjan), Iran.

## Funding

This work was supported by department of soil science, Islamic Azad University, Azerbayjan-West branch.

## Acknowledgement

We would like to express our special thanks and gratitude to department of agriculture in Azerbayjan-west University that provided such an amazing opportunity for us to work in their field and supported us financially.

## Declaration of Generative AI and AI-Assisted Technologies in the Writing Process

During the preparation of this work, the author(s) did not use any AI and/or AI-assisted technologies and authors take full responsibility for the content of the publication.